# ENERGY PRICE AND WORKLOAD RELATED DISPATCHING RULE: BALANCING ENERGY AND PRODUCTION LOGISTICS COSTS

Balwin Bokor[1], Wolfgang Seiringer[1], Altendorfer Klaus[1], and Felberbauer Thomas[2]

[1]Dept. of Production Operation Management, University of Applied Sciences Upper Austria,
Steyr, Upper Austria, OOE, AUSTRIA
[2]Dept. of Media and Digital Technologies, St. Pölten University of Applied Sciences,
St. Pölten, Lower Austria, NOE, AUSTRIA

**ABSTRACT**

In response to the escalating need for sustainable manufacturing practices amid fluctuating energy prices, this study introduces a novel dispatching rule that integrates energy price and workload considerations with Material Requirement Planning (MRP) to optimize production logistics and energy costs. The dispatching rule effectively adjusts machine operational states, i.e. turn the machine on or off, based on current energy prices and workload. By developing a stochastic multi-item multi-stage job shop simulation model, this research evaluates the performance of the dispatching rule through a comprehensive full-factorial simulation. Findings indicate a significant enhancement in shop floor decision-making through reduced overall costs. Moreover, the analysis of the Pareto front reveals trade-offs between minimizing energy and production logistics costs, aiding decision-makers in selecting optimal configurations.

## 1 INTRODUCTION

Historically, production companies have focused on timely delivery, cost efficiency, and product quality. However, in recent years, the environmental impact of production has become increasingly significant. Energy costs have emerged as a crucial factor in both economic and environmental terms. Previously underrepresented in operations research literature, the integration of dynamic energy cost rates into operational planning has gained prominence due to the global energy crisis that peaked between 2022 and 2023 (Skėrė et al. 2023). This crisis has highlighted the volatility of energy prices within the European Union, where prices saw significant increases throughout. Efforts to diversify energy sources have been accelerated, yet European countries continue to experience high price variability influenced by geopolitical and market factors. Furthermore, the push for sustainability has prompted numerous companies to reduce their dependence on fossil fuels, as seen in various European countries. This trend is underscored by insights from the The World Economic Forum (2023), as noted by Singh et al. (2019). Amidst these challenges, production companies are increasingly prioritizing environmental concerns, emphasizing sustainable production methods and the integration of renewable energy technologies into their operations. This focus is addressed by Hegab et al. (2023). Hence, operational strategies need to incorporate energy decisions that align with sustainability objectives, a process each company must undertake individually (Taghavi 2022).

Therefore, we investigate the integration of dynamic energy prices into operations research, acknowledging the growing need for research in this area for both academic and practical applications. Our focus lies in exploring the evolving role of energy cost decisions in production dispatching, highlighting the dual imperatives of cost efficiency and environmental sustainability. To this end, we develop an energy price and workload related dispatching rule and combine it with a medium-term production planning and control system, in particular Material Requirement Planning (MRP). Thus, MRP is employed to establish and schedule production orders, while our dispatching rule is utilized for shop floor processing. This dispatching rule determines the operational state of machines, i.e. on or off, based on the current energy price and workload. Machines remain off if energy prices surpass a predefined threshold and workload allows a pause without risking delays that could elevate operational costs as tardiness increases. To evaluate the performance of our dispatching rule, we integrate it into a stochastic multi-item multi-stage job shop production system. Our performance evaluation employs a full factorial simulation study, assessing overall costs encompassing production logistics costs such as work in progress (WIP) and finished goods inventory (FGI), alongside with energy costs. Thereby, we explore the interplay between production logistics and energy costs by identifying the Pareto front. Additionally, we examine the effects of different MRP planning parameters and dispatching rule parameters on overall costs and production system behavior. Hence, following research questions are addressed:

- *RQ1: How can decisions related to energy prices be integrated into dispatching and combined with medium-term production planning and control systems?*
- *RQ2: What impact do energy price and workload related dispatching decisions have on the behavior of the production system, particularly on the operational state of machines?*
- *RQ3: How does an energy price and workload related dispatching rule affect the overall costs, and what are the trade-offs between energy costs and production logistics costs, i.e. Pareto front?*

This research provides important insights for both academic research and managerial practice. Academically, it contributes by introducing a novel dispatching rule related to energy price and workload and integrating it into medium-term production planning and control systems. Managerially, it offers decision-makers a method to balance energy costs and production logistics, along with an analysis of their interplay through the Pareto frontier.

This publication is organized as follows: Section 2 offers a concise overview of the literature on energy reduction in manufacturing, focusing on process control that involves changing the operational states of machines. Section 3 introduce our developed energy price and workload related dispatching rule. This is followed by an explanation of the simulation model and conducted the numerical study in Section 4. Afterwards, Section 6 discusses the results. The publication concludes with final thoughts and suggestions for further research.

## 2 LITERATURE REVIEW

A comprehensive overview of energy aspects in production simulation comes from Peter and Wenzel (2017) and Wenzel et al. (2018). Peter and Wenzel (2017) presented different methods to consider energy related aspects in simulation. Whereas Wenzel et al. (2018) summarized different simulation modelling methods related to energy and emissions. Moreover, the authors aggregated common objectives, requirements and implementation procedures of simulation studies. Complementing these findings, a literature review by Renna and Materi (2021) highlights the substantial energy consumption and greenhouse gas production associated with manufacturing systems. They emphasize the need for more focused research on energy-efficient and sustainable production planning, summarizing key papers on energy efficiency and renewable energy sources.

Energy reduction in manufacturing extends from individual device-level strategies to comprehensive enterprise-level approaches that include supply chain operations, as noted by in Reich-Weiser et al. (2010). Duflou et al. (2012) outlined three primary strategies at the device level: 1) the optimization of machine tool design, 2) enhancement of process control, and 3) process/machine tool selection. Optimized machine tool design boosts efficiency by integrating technological advances, recovering waste, and streamlining operational inputs such as compressed air. Enhanced process control aims to lower energy by implementing selective device or machine shutdowns, reducing idle machine times, optimizing processing parameters, i.e. temperature, pressure, flow rates, or energy efficient process modeling and planning. Whereas, energy reduction through process/machine tool selection is accomplished by selecting alternative production resources as discussed by Li et al. (2016) or by optimizing machine tool capacity. The process control to reduce energy consumption includes notable contributions from Frigerio and Matta (2016), who address stochasticity in production and strive to establish energy-efficient switching of machines. In their earlier work Frigerio and Matta (2014) enhanced energy efficiency by refining a threshold policy to reduce machine energy consumption during idle periods. This method incorporates warm-up time modeling and uses experimental data to determine optimal parameters. Further advancing the discussion on energy efficiency, Frigerio et al. (2024) developed and analyzed buffer-based threshold policies for controlling multiple machines simultaneously in a serial production line. Their optimal control policies are designed to minimize energy consumption while achieving a specific production rate, using discrete event simulation to evaluate various machine group clustering and buffer-threshold settings. Their findings underscore the importance of careful selection and control of machines and thresholds to optimize performance. Similarly, Loffredo et al. (2024) utilized buffer information to optimize machine transitions between active and standby states, which demand less energy. With stochastic startup times, they aimed to minimize energy use within production constraints. Using a Markov Decision Process and linear programming, they efficiently managed a two-stage system and extended this approach to more complex systems.

These studies collectively highlight the evolving methods and the critical importance of integrating energy-efficient decisions across various levels of manufacturing, from individual machine adjustments to broad-scale operational modifications. This holistic approach is crucial for significantly reducing the industry's environmental footprint and moving towards more sustainable manufacturing practices.

## 3   ENERGY PRICE AND WORKLOAD RELATED DISPATCHING RULE

To optimize the balance between energy costs and potential tardiness and the related inventory costs, we have developed a machine related dispatching rule that takes into account both the current energy price and the workload of the machine. The following Figure 1. visualizes the key elements of the developed dispatching rule.

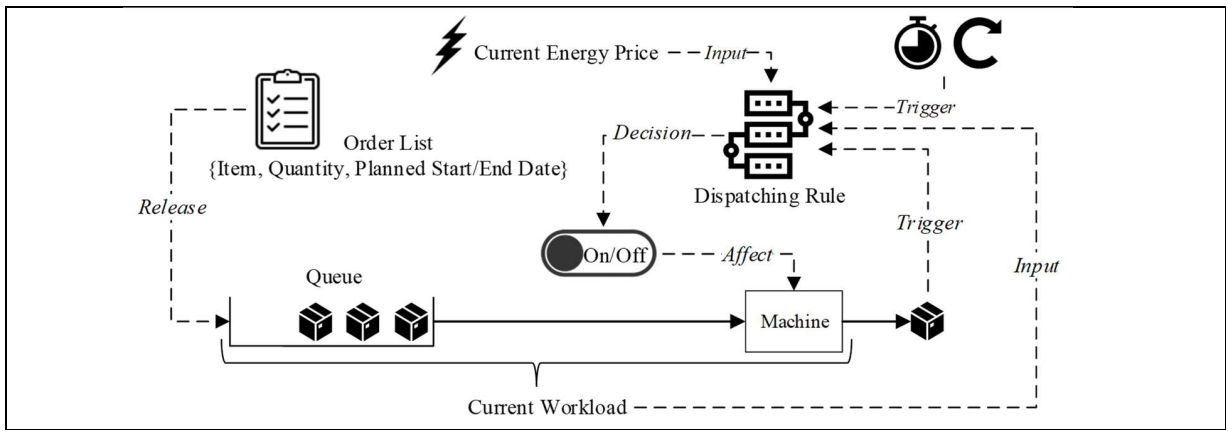

Figure 1: Abstracted Logic Dispatching Rule.

As shown in Figure 1, an order list is generated by a production planning and control system, such as Material Requirement Planning (MRP). The conceptual explanation of MRP is provided by Hopp and Spearman (2011). This order list includes all production orders with details on the item, quantity as well as planned start and end dates. These production orders are released onto the shop floor when input material is available, and the planned start date is reached. At the shop floor, the production orders await processing. During processing, our developed dispatching rule, which considers energy prices and workload, dictates the machine's operational state, i.e., whether to turn the machine on or off. As illustrated in Figure 1, this decision-making process is triggered for each machine whenever a production order is in the queue, either after a production order has been processed or after a predetermined time period, i.e. occurring periodically. To make decision on the machine's operational state, our dispatching rule evaluates two inputs: the current energy price and the machine's current workload, both of which are compared against calculated thresholds. To provide a detailed explanation of our developed dispatching rule, the following Figure 2 visualizes the rule in a pseudo-code format.

```
Algorithm 1 Energy Price and Workload Related Dispatching Rule
 1: while simulation model is running do
 2:     wait for the next event:
 3:         - next production order is processed
 4:         - time period elapses
 5:     call MACHINEOPERATIONALSTATE(currentSimTime, machine)
 6: end while
 7: function MACHINEOPERATIONALSTATE(currentSimTime, machine)
 8:     currentEnergyPrice ← getCurrentEnergyPrice(currentSimTime)
 9:     energyThreshold ← averageLongTermEnergyPrice × energyPriceFactor
10:     machineOn ← false
11:     if (currentEnergyPrice < energyThreshold) then
12:         machineOn ← true
13:     else
14:         currentWorkload ← 0
15:         for each order in machineQueue do
16:             currentWorkload += processingTime + setUpTime
17:         end for
18:         workloadThreshold ← availableDailyCapacity × capacityFactor
19:         if (currentWorkload > workloadThreshold) then
20:             machineOn ← true
21:         else
22:             machineOn ← false
23:         end if
24:     end if
25:     return machineOn
26: end function
```

Figure 2: Pseudo-Code Dispatching Rule.

As seen in Figure 2 *(row 11)* and *(row 19)* the workload related dispatching rule performs two comparisons. Initially, as indicated in *(row 11)*, the rule compares based on the current energy price, which is queried at the moment of decision-making. Consequently, as described in Section *1 Introduction*, short-term fluctuations in energy prices within a day are considered. For example, energy is typically more available during daylight hours and summer months due to increased solar output. This queried energy price is then compared to an energy threshold. Whereby the energy threshold is calculated by taking the average energy price for the month and multiplying it by an energy factor *(row 9)*, which is enumerated in our numerical study. If the current energy price is below this energy threshold, the machine is turned on *(row 12)*, as the price is favorably low. However, as seen in Figure 2 *(row 13)* if the current energy price exceeds the energy threshold, a workload comparison is performed *(row 19)*. For the workload comparison, the current workload for each machine is calculated by cumulating the product of all waiting production orders, i.e. in the queue, with their respective expected mean planned process times and expected mean set-up times *(row 15-17)*. The workload increases as production orders arrive at the machine, i.e. waiting in queue, and decreases after processing. For example, if three production orders are pending, each with an expected mean planned process time of 2 time units and an expected mean set-up time of 1 time unit, the current workload of the machine totals 9 time units. This workload is compared to a workload threshold *(row 19)*. Whereby the workload threshold is calculated by multiplying the available daily capacity with a capacity factor *(row 18)*, which is also enumerated in our numerical study. If the current workload exceeds this workload threshold, the machine is turned on *(row 20)* to handle the high workload and mitigate the risk of potential tardiness. Conversely, if the workload is below the workload threshold the machine remains off *(row 22)*, as the current energy price is high, and the low workload allows for deferred processing with minimal risk of tardiness.

## 4 SIMULATION MODEL

To assess the performance of our energy price and workload related dispatching rule, we developed a stochastic multi-item multi-stage simulation model and integrated the algorithm into it. The production system under study is modeled using an agent-based discrete-event approach in AnyLogic software version 8.8.6. The developed simulation model is generic, allowing the simulation of various production system structures, demand scenarios, and energy price scenarios through the appropriate parameterization of the Bill of Materials (BoM), work plans, and energy prices. In this publication, we focus on a job shop production system structure. We first outline the production system, including its structure, the BoM, the processing of production orders, and customer demand. Subsequently, we delve into production planning and dispatching, by detailing the applied medium-term production planning and control system and the integration of our dispatching rule. This builds upon the foundational concepts presented in Section 3 *Energy Price and Workload Related Dispatching Rule*. Lastly, we discuss the comprehensive numerical study.

### 4.1 Production System

We explore a stochastic multi-item multi-stage job shop production system with 8 items {101, 102, 103, 104, 105, 106, 107, 108} and 4 machines {M1.1, M1.2, M1.3, M1.4}. The production system structure, including the BoM and the routing, is illustrated in Figure 3. As shown, the BoM consists of two levels: level 0, which includes the eight items, and level 1, which comprises component 202. This component is always available and does not require planning. The processing sequence, related to individual items, is marked with green-bordered letters starting from 'A' onward. For instance, item 101 undergoes processing first on M1.1 (A), then on M1.3 (B), followed by M1.2 (C), and concludes on M1.4 (D).

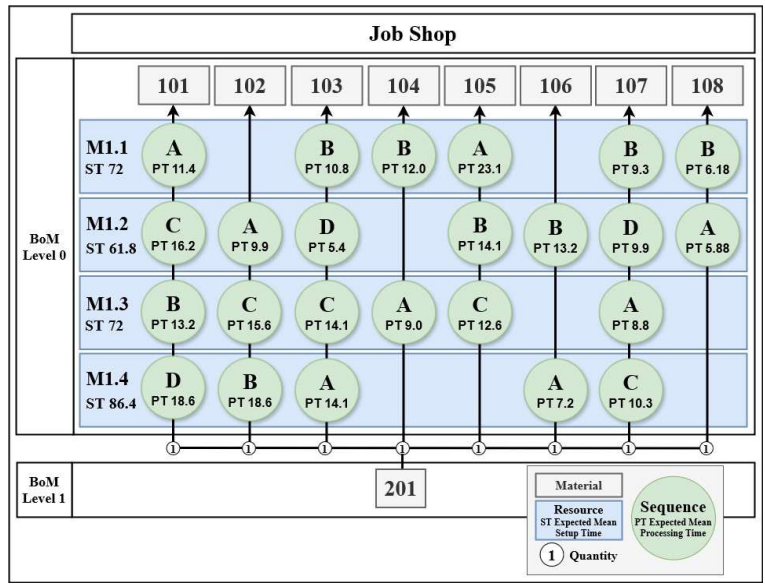

Figure 3: Investigated Production System Structure including Bill of Material.

The available capacity at each machine is 1440 time units per day. For a systematic examination of the observed production system and to address the stated research questions through our simulation study, we established a planned utilization of 80%. To achieve this planned utilization, the required capacity per machine per day is set at 1024 minutes. Since the ratio of available and required capacity is identical across all machines, no bottleneck exist. The expected mean processing time varies for each item and machine. Whereas, the expected mean setup time is consistent within each machine, accounting for 10% of the available capacity. This results in decreased setup times as the number of items or components processed on a single machine increase, and vice versa. Both the expected mean processing times and expected mean setup times are detailed in Figure 3. Each item follows a set processing sequence that includes a setup phase before operation on each machine, incorporated into the simulation as setup time. This setup time is applied even when an identical item follows on the same machine. To establish the required capacity for the production system, customer orders containing only one of the eight item types are generated. The interarrival time for these customer orders follows a lognormal distribution, with an expected mean of 10 periods for each item and a coefficient of variation (CV) of 0.2. This means that, on average, a customer order for each item is generated every ten periods. To incorporate additional demand uncertainty, the customer order quantity also varies according to a lognormal distribution, with each item having a different expected mean quantity and a CV of 0.5. Additionally, the customer-required lead time includes a fixed part of 10 periods and a variable part that follows a lognormal distribution with an expected mean of 5 periods and a CV of 0.5.

## 4.2 Production Planning and Dispatching

To process the customer orders MRP is selected as medium-term production planning and control system. Orlicky (1975) developed MRP, a push-type production and inventory control system that operates through four centrally controlled steps: netting, lot-sizing, backward scheduling, and BOM explosion, as detailed by by Hopp and Spearman (2011). Specified for this publication we apply FOP lot sizing policy and the MRP structure is similar to Seiringer et al. (2023). The MRP run is performed each period until the end of the simulation. Based on MRP, an order list is generated that includes all production orders with details on the item, quantity, as well as planned start and end dates, as illustrated in Figure 1. It is important to note that our developed dispatching rule is generic in nature and can be applied across various production planning and control systems. The generated production orders are released to the shop floor when the planned start date is reached and after successfully verifying material availability. At dispatching, our energy price and workload related rule determines the machine's operational state, i.e. turning the machine on or off. To facilitate this, we have integrated our developed algorithm, which is conceptually visualized in pseudo-code format in Figure 2, into the shop floor processing agent. The subsequent Figure 4 visualizes the shop floor processing agent in the simulation model.

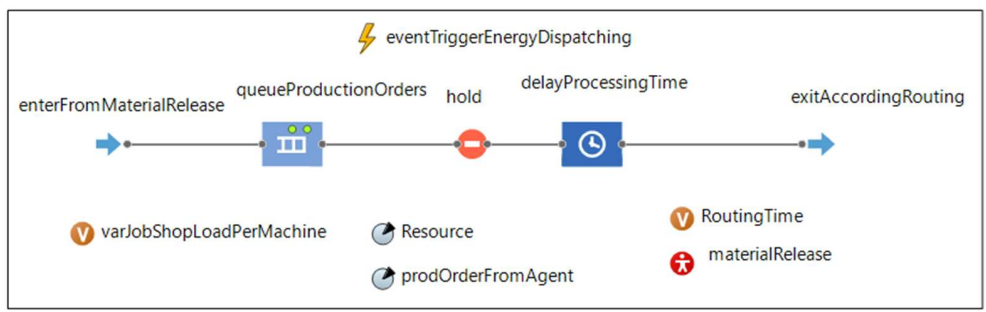

Figure 4: Shop Floor Processing Agent in Simulation Model.

Instead of using individual agents for each machine, a general implementation processes orders by pulling the parameterized production system structure from a relational database and initializing an agent for each machine. Production orders are then sequentially passed from one machine to the next according to the established routing sequence. As illustrated in Figure 4, if the planned start date is reached and the required materials are available, the production order is transferred to the *"enterFromMaterialRelease"* function of the machine agent, which handles shop floor processing. The processing of the production order is modeled by the *"delayProcessingTime"*. The machine's operational state is modeled using the *"hold"* function. At dispatching the machine's operational state is determined. Decision-making regarding the machine's operational state occurs only when a production order is in the *"queueProductionOrder"* and awaiting processing. This decision-making is triggered either after a production order has been processed or periodically, i.e. every hour, by the event "*eventTriggerEnergyDispatching*".

As stated in Section 3 *Energy Price and Workload Related Dispatching Rule* the machines operation state is determined based on two comparisons. At first, the current energy price, which is queried at the moment of decision-making, is compared to an energy threshold. This threshold is computed by taking the average energy price for the month and multiplying it by an energy factor. The following Figure 5 a) illustrates the short-term energy price per Megawatt hour (MWh), i.e. the energy price for each hour within a day, for the month of June, as well as the average energy price per MWh for this month. Whereas Figure 5 b) shows, the average energy price for each month. These energy prices are derived from actual energy price data for Austria in 2023 and can be accessed at online platform SMARD.

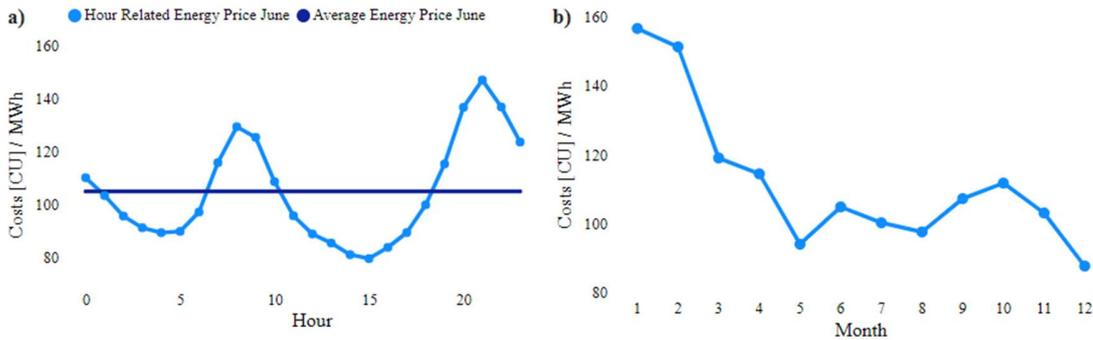

Figure 5: (a) Short-Term and Average Energy Price June and (b) Average Monthly Energy Price.

As illustrated on Figure 5 a), energy prices in June, particularly between 13:00 and 16:00 hours, are low due to increased solar output. Conversely, as shown on Figure 5 b), energy prices are especially high in the first months of the year because Austria, being in a temperate climate zone, experiences cold winters that require more energy consumption. Whenever the current energy price exceeds the energy threshold, a secondary comparison based on workload is conducted. For example, if the energy price factor is set to 1, this secondary comparison occurs when the current energy price is higher than the average for the month. For the workload comparison, the workload threshold is calculated by multiplying the available daily capacity of 1440 minutes by a capacity factor. As outlined in Section 3 *Energy Price and Workload Related Dispatching Rule*, if the current workload surpasses this threshold, the machine is activated to manage the increased workload and mitigate the risk of potential tardiness. Otherwise, the machine is turned off.

### 4.3 Numerical Study

We conduct a comprehensive numerical study to evaluate the performance and observe the behavior of our developed energy price and workload related dispatching rule. Therefore, we perform a full factorial simulation study, enumerating both the parameters of our dispatching rule, i.e., the energy factor and the capacity factor, and the MRP planning parameters, i.e., lot size, safety stock, and planned lead time. Various parameters for the dispatching rule are tested, as they determine the behavior of the rule through the calculation of energy and workload thresholds, as depicted in Figure 2. Meanwhile, different MRP planning parameters are tested because they significantly influence the production system performance, as shown by Seiringer, Altendorfer et al. (2022) and to assess their interplay with the dispatching rule. Although MRP planning parameters are set on an item level, we employ identical planning parameters across all items to mitigate the combinatorial effect. Concerning MRP planning parameters: we set the planned lead time in days; apply FOP lot-sizing policy measured in days; and safety stock levels are established as a proportion of the item's expected customer order quantity. For example, setting a safety stock level 2 with an expected customer order quantity of 50 results in a total safety stock of 100. The following *Table 1* visualizes the tested parameter ranges, which were established based on preliminary simulation runs, for our full factorial enumeration.

Table 1: Investigated MRP and Dispatching Rule Parameters.

|  | Min | Max | Step Size | Iterations |
|---|---|---|---|---|
| *MRP Planning Parameters:* | | | | |
| Planned Lead Time Items [days] | 1 | 10 | 1 | 10 |
| Safety Stock Items [prop. order quantity] | 0 | 2.0 | 0.5 | 5 |
| Lot Size Items [days] | 1 | 6 | 1 | 6 |
| *Dispatching Rule Parameters:* | | | | |
| Energy Factor | 0.50 | 1.40 | 0.10 | 10 |
| Capacity Factor | 0.25 | 2.50 | 0.25 | 10 |
| Total iterations | | | | 30,000 |
| Total simulation runs with 10 replications | | | | 300,000 |

In As depicted in *Table 1*, a total of 30,000 iterations were conducted. The observed production system introduces stochastic elements through demand and processing, elaborated in Section *4.1 Production System*. To account for these stochastic influences on simulation results, each iteration was replicated 10 times, resulting in 300,000 individual simulation runs. Each replication lasts 400 days, with a warmup phase of 150 days, after which the simulation environment statistics are reset. To streamline the simulation process, parallel computing is employed, distributing the study across 21 computers to reduce simulation time. Parameter combinations are generated using RStudio, uploaded to a PostGreSql database, and subsequently queried by the AnyLogic simulation model. Finally, the results are recorded back into the PostGreSql database for analysis.

## 5 NUMERICAL RESULTS

The energy price and workload related dispatching rule is evaluated based on overall costs, encompassing production logistics costs such as WIP, FGI, and tardiness costs, along with energy costs. The production logistics cost structure includes 1 cost unit (CU) per day for WIP, 2 CU per day for FGI, and 38 CU per item per day for tardiness. This cost structure aims to achieve a 95% service level, reflecting the strategic emphasis on reducing tardiness. Additionally, finished goods storage costs are set at twice that of the WIP to reflect higher expenses due to tied-up capital. Concerning energy costs, in contrast to the consideration of machine energy consumption over time as in Bokor and Altendorfer (2023), this study assumes a fixed energy requirement of 1 kilowatt-hour (kWh). Energy costs are calculated by multiplying this fixed energy requirement by the realized processing time and the realized energy price upon machine entry. The study does not consider the warm-up time between machine start and machine stop. This could be very relevant for domains especially where the material temperature is relevant for the production process.

### 5.1 Impact of Planning Parameters

As introduced above, the MRP planning parameters, i.e. lot size, planned lead time and safety stock, have been enumerated in combination with the developed dispatching rule parameters, i.e. energy factor and capacity factor. Literature, as exemplified by Altendorfer (2019) or Seiringer, Castaneda et al. (2022), demonstrates the significant impact of planning parameters on production logistics costs. Given our focus on both production logistics and energy costs, this section examines the influence of planning parameters on overall costs and the interplay between optimized planning parameters. The analysis reveals that in most cases, a safety stock of 0 is optimal. Thus, the results omit specific safety stock outcomes, consistently applying the best safety stock parameter. The following Figure 6 shows the overall costs per day related to planned lead time and lot size, always in combination with the respective other best parameter. For example, Figure 6 a) shows that a planned lead time of 5 days results in the lowest costs when combined with a lot size of 1. Detailed examination of the results reveals that, for the studied production system, costs exhibit convex behavior concerning the planned lead time, consistent with findings by Altendorfer (2019) and Seiringer, Castaneda et al. (2022). The introduction of additional energy costs does not alter this fundamental relationship. However, the results related to lot sizing effects are intriguing, showing cost escalation with increasing lot size, contrary to the typically reported convex relationship in literature as discussed in Altendorfer (2019) or Seiringer et al. (2023). We propose that the inclusion of energy costs favors lower lot sizes, enabling the system to leverage varying energy prices throughout the day when employing the energy related dispatching rule. It's worth noting that these results are limited to analysis of a single production system structure, and broader generalization would necessitate examination across various production system structures and planned shop loads. Moreover, consistent with existing literature, an intuitive observation is that higher lot sizes result in longer planned lead times, as shown in Figure 6 b).

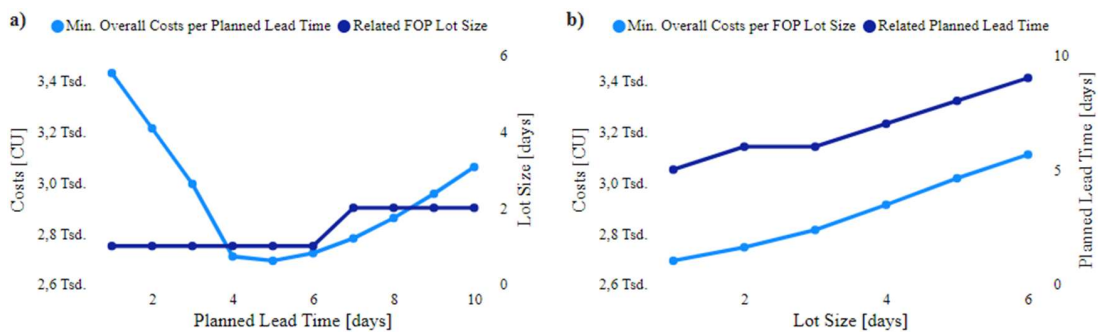

Figure 6: Overall Cost per Day in Relation to (a) Planned Lead Time and (b) FOP Lot Size

## 5.2  Impact of Dispatching Rule Parameters

This paper introduces a dispatching rule which is parametrized based on energy factor and the capacity factor. Both influence different aspects of production system performance. The energy factor determines the acceptable level of energy prices for continuing production, while the capacity factor addresses the amount of workload manageable before halting production due to high energy prices. Figure 7 illustrates the effects of these factors on both overall and energy costs per day.

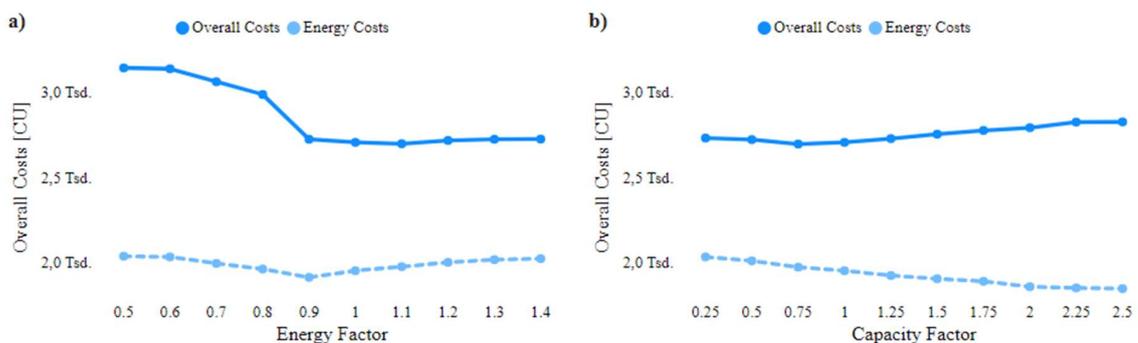

Figure 7: Overall / Energy Costs per Day in Relation to (a) Energy Factor and (b) Capacity Factor

In detail Figure 7 a) shows the relationship between varying the energy factor (with the capacity factor held constant at 1) and both overall and energy costs. The relationship appears convex, with the optimal energy factor identified at 0.9. Interestingly, energy costs increase for factors below 0.9, suggesting more frequent machine stoppages due to increased sensitivity to energy prices. Figure 7 b) shows a similar convex trend of the overall costs when the capacity factor is varied (with the energy factor held constant at 1). However, a rise in the capacity factor consistently lowers energy costs, reflecting higher workload acceptance despite increased energy prices. Nonetheless, increasing the capacity factor beyond 1 results in a disproportionate rise in production logistics costs relative to the decrease in energy costs, leading to higher overall costs. This demonstrates a complex trade-off between energy costs and production logistics costs.

To delve deeper into the behavior of the developed dispatching rule, Figure 8 presents the number of occasions when energy price exceeds the calculated energy threshold, prompting a decision to switch off the machine, and the instances when the machine is actually turned off, depending on the workload. These findings are based on the same parameters as those in Figure 7. In detail Figure 8 a) reveals that a lower energy factors result in more instances where the machine should ideally be turned off, but the interaction with the workload reduces the actual number of machine turn-offs for energy factors below 0.8. This interaction explains the observed energy cost effects. Conversely, Figure 8 b) shows that the pattern of machine turn-offs displays a concave relationship with the capacity factor, attributed to the optimal planning parameters derived from MRP. Under constant conditions, this value would typically remain steady. However, an increase in actual machine turn-offs as the capacity factor rises from 0.25 to 1.75 confirms the model's validity. Notably, a decrease in machine turn-offs at a capacity factor of 2, followed by a slight increase, further illustrates the complex interaction between the energy-focused dispatching rule and the production planning parameters.

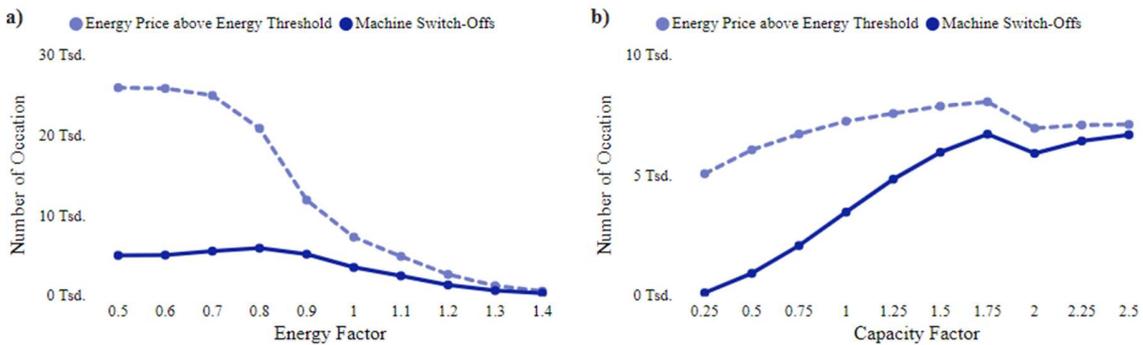

Figure 8: Number of Machine Switch-Offs due to (a) Energy Factor and (b) Capacity Factor

### 5.3 Pareto Optimality

Since the focus of this study is on the trade-off between production logistics costs and energy costs, Figure 9 presents the solution points. Each solution point corresponds to a specific set of planning and dispatching parameters and the respective Pareto front for the simulation study. To improve the clarity of the Figure 9, all production logistics costs above 8,000 CU have been omitted. The results illustrate the balancing act between energy and production logistics costs, indicating a tendency towards accepting higher energy costs at the global optimum. Additionally, the results offer the option to choose higher production logistics costs in favor of lower energy costs, given the difficulty in precisely quantifying tardiness costs within real production systems. As depicted in Figure 9, the optimal overall costs are achieved at 1,494 CU for energy and 743 CU for production logistics, resulting in a total of 2,692 CU in overall costs. Overall, the findings demonstrate that the developed dispatching rule, in conjunction with the well-known MRP planning method, functions effectively and can be tailored to prioritize either energy or production logistics costs.

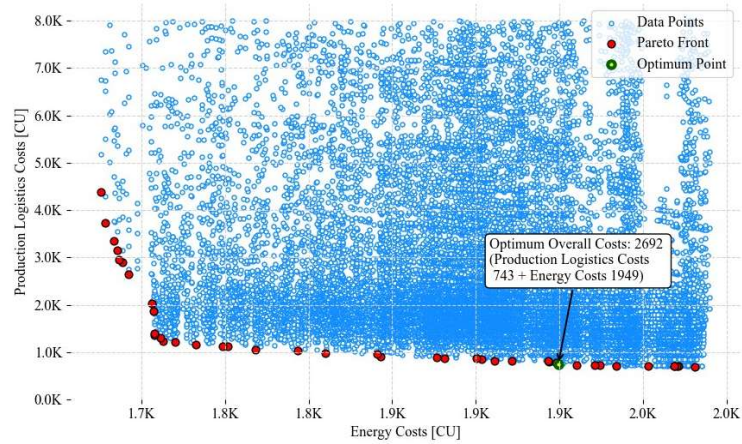

Figure 9: Pareto Front of Energy and Production Logistics Costs.

## 6 CONCLUSION

In this publication, we develop an energy price and workload related dispatching rule and combine it with MRP to improve both energy costs and production logistics costs. Our developed dispatching rule adjusts the machine's operational state, i.e., whether to turn the machine on or off, based on the current energy price and the machine's current workload, both of which are compared against calculated thresholds. To assess the performance of our dispatching rule, we develop a stochastic multi-item multi-stage job shop simulation model and integrated the algorithm into it. We conduct a comprehensive full-factorial simulation study to evaluate the performance and observe the behavior of our dispatching rule. Thereby, we enumerate both the parameters of our dispatching rule, i.e., the energy factor and the capacity factor, and the MRP planning parameters, i.e., lot size, safety stock, and planned lead time. Our findings reveal that optimal settings for the MRP planning parameters, particularly a lower lot size and minimal safety stock, significantly reduce both production logistics and energy costs. The dispatching rule further demonstrates its effectiveness by dynamically adjusting the machine's operational state based on energy price fluctuations and workload, leading to improved operational decisions. The interplay between energy and capacity factors underscores the nuanced balance required to minimize costs while maintaining production efficiency. Notably, the analysis of the Pareto front highlights the trade-offs involved, demonstrating that while some configurations favor lower energy costs, others optimize production logistics costs, enabling decision-makers to select configurations that best align with their operational priorities and constraints. Future research could explore this dispatching rule across diverse production environments and refining the models to incorporate real-time energy price signals, a warm-up time, or the increase of energy consumption between machine turn off and turn on. Additionally, integrating forecasts of future energy prices or workloads, as well as dynamic capacity setting methods, could enhance the algorithm's functionality and efficacy.

## ACKNOWLEDGMENTS

This project is financed by research subsidies granted by the government of Upper Austria [BF CoE-E/SP] and the PhD-funding program of the University of Applied Sciences Upper Austria EE-Scheduling [FO999905125], funded by the Austrian Research Promotion Agency FFG.

## AUTHOR BIOGRAPHIES

**WOLFGANG SEIRINGER** is a Research Associate in the field of Operations Managemen at the University of Applied Sciences Upper Austria and Ph.D. candidate at the JKU Linz for Logistics and Supply Management. Email: wolfgang.seiringer@fh-steyr.at.

**BALWIN BOKOR** is a Research Associate in the field of Operations Management at the University of Applied Sciences Upper Austria and Ph.D. candidate at the JKU Linz for Logistics and Supply Management. Email: balwin.bokor@fh-steyr.at.

**KLAUS ALTENDORFER** is a Professor of Operations Management at the University of Applied Sciences Upper Austria, expert in production system simulation, stochastic inventory model, and production planning. Email: klaus.altendorfer@fh-steyr.at.

**THOMAS FELBERBAUER** is Academic Director of the Smart Engineering program at St. Pölten University of Applied Sciences, expert in production planning and simulation. Email:thomas.felberbauer@fhstp.ac.at.